\documentclass[aps, prd, twocolumn, amsmath, amssymb, floats, floatfix, groupedaddress, nofootinbib, nopacs,notitlepage]{revtex4-1}

\usepackage{latexsym}
\usepackage{multirow}
\usepackage{algpseudocode}
\usepackage{algorithm}
\usepackage{graphicx}
\usepackage{amsmath}
\usepackage{amsfonts}
\usepackage{mathrsfs}
\usepackage{mathtools}
\usepackage{amssymb}
\usepackage{times}
\usepackage{xspace} 
\usepackage[usenames]{color}
\usepackage{dcolumn}
\usepackage{bm}
\usepackage{mathrsfs}
\usepackage{tikz}
\usepackage{appendix}
\usepackage{dsfont}
\usepackage{breqn}
\usepackage[]{units}       
\usepackage{epstopdf}
\usepackage{breqn}

\usepackage{subfigure}		
\usepackage[normalem]{ulem} 

\newcommand{\deri}[2]{\dfrac{\textnormal{d} #1 }{\textnormal{d} #2}}
\newcommand{\iu}{{i\mkern1mu}}

\usepackage[tracking=true]{microtype}
\SetTracking{}{500}
\SetTracking{encoding={*}, shape=sc}{40}

\makeatletter
\let\cat@comma@active\@empty
\makeatother
\begin{document} 
 
\title{Electromagnetic quasinormal modes of Schwarzschild--anti--de Sitter black holes: \\ Bifurcations, spectral similarity, and exact solutions in the large black hole limit}

\author{Sean Fortuna}
\email{sjfortuna@nip.upd.edu.ph}
\author{Ian Vega}
\email{ivega@nip.upd.edu.ph}
\affiliation{National Institute of Physics, University of the Philippines, Diliman, Quezon City 1101, Philippines}
\date{\today}

\begin{abstract}
We revisit the peculiar electromagnetic quasinormal mode spectrum of an asymptotically anti--de Sitter Schwarzschild black hole. Recent numerical calculations have shown that some quasinormal mode frequencies become purely overdamped at some critical black hole sizes, where the spectrum also bifurcates. In this paper, we shed light on unnoticed and unexplained properties of this spectrum by exploiting some novel analytic results for the large black hole limit. We demonstrate, both numerically and analytically, that the quasinormal mode spectra of large black holes become approximately isospectral, and refer to this new symmetry property as spectral similarity. We take advantage of this spectral similarity to derive a precise analytic expression for the locations of the bifurcations, in which a surprising Feigenbaum-like constant appears. We derive an exact solution for its spectrum and eigenfunctions, and find that large black holes cannot be made to vibrate with electromagnetic perturbations, independently of the boundary conditions imposed at spatial infinity. Finally, we characterize the insensitivity of the spectrum to different boundary conditions by analyzing the expansion of the quasinormal mode spectrum around the large black hole limit.
\end{abstract}

\maketitle

\section{Introduction}

The study of the quasinormal mode spectrum of asymptotically anti--de Sitter (AdS) spacetimes is heavily motivated by the AdS/CFT correspondence \cite{son2002, kovtun2005, berti2009, hubeny2015}. The quasinormal mode spectrum of an AdS black hole may offer information on the dual conformal field theory (CFT), such as the stability of a thermal dual conformal theory and its relaxation times or transport properties of its dual conformal fluid \cite{rama1999, danielsson2000, friess2007, varghese2009}.  

The Schwarzschild--AdS (SAdS) black hole is a well-trodden path. Historically, the first investigation of its quasinormal mode spectrum was conducted in \cite{chan1997}, while its implications in CFT were first described in \cite{horowitz2000}. Since then, a wealth of papers has been written for different dimensions, coupled perturbing fields, boundary conditions, as well as using different numerical and analytical methods \cite{cardoso2001,cardoso2003,berti2003,cardoso2004,giammatteo2005,musiri2003c,musiri2006,
michalogiorgakis2007,myung2008,skenderis2009,bakas2009,bakas2009a, wang2021,wang2021a}.

The vibrational modes of linear perturbations of a black hole spacetime is determined by the spacetime's properties. For the SAdS black hole, these are the black hole radius $\bar{r}_h$ and the AdS radius $\bar{L}$. Together with appropriate boundary conditions on the black hole horizon and at spatial infinity, these modes form a discrete spectrum listed, for spherically symmetric spacetimes, by two numbers: its overtone number $n$ and its harmonic mode index $l$.

In this paper, we revisit electromagnetic (EM) perturbations of the SAdS black hole, concentrating on the large black hole limit, which is defined as
\begin{equation}
\rho := \dfrac{\bar{L}}{\bar{r}_h}, \qquad \rho \ll 1.
\end{equation}
Compared to its scalar and gravitational counterparts, the electromagnetic quasinormal mode spectrum of a SAdS black hole has been shown to be rather peculiar. Here is a list of observations and open questions.
\begin{enumerate}
\item As the dimensionless parameter $\rho$ is varied, a bifurcation in the spectrum occurs \cite{wang2021,wang2021a}. This happens whenever a pair of quasinormal modes (QNM) satisfying
\begin{equation}
\bar{\omega} = - \bar{\omega}^*
\label{eq:symmetry}
\end{equation}
meet up at the negative imaginary axis, where they split into two purely imaginary or overdamped modes. What is the character of this bifurcation, and can we predict when it occurs?
\item As the dimensionless parameter $\rho$ approaches zero, these overdamped modes populate the lowest lying eigenfrequencies \cite{cardoso2001}. In the large black hole limit, can the spacetime still vibrate? \cite{cardoso2003}
\item There is a freedom in choosing the boundary conditions (BC) at spatial infinity, which may be motivated by the ease of calculation thereafter or by considerations from the AdS/CFT correspondence \cite{son2002, moss2002,michalogiorgakis2007,skenderis2009,bakas2009,bakas2009a}. However, there seems to be some evidence that the spectrum in the large black hole limit is insensitive to this choice of BC (either Dirichlet or Robin) whenever a QNM spectrum results \cite{moss2002}.
\item In the large black hole limit, the spectrum is strongly independent of the harmonic mode index $\ell$ even though the potential is strongly dependent on $\ell$ \cite{cardoso2001,cardoso2003,cardoso2004}. 
\end{enumerate}
In this paper, we expound and answer these questions.

First, we shall show that there is a fundamental difference in the asymptotic behavior of bifurcated and unbifurcated modes. As we shall see, with the appropriate rescaling, the unbifurcated modes approach a level spacing of
\begin{equation}
\dfrac{\sqrt{3}}{2} - \dfrac{3 \iu}{2} = 0.866025 - 1.5 \iu,
\end{equation}
in the large overtone and large black hole limit \cite{cardoso2004}, while the level spacing of bifurcated modes approach $- \iu.$

We also conjecture a numerically motivated Feigenbaum-like constant characterizing the bifurcation, and derive a master equation that accurately approximates the value of $\rho$ when the bifurcations occur.

Second, we shall derive the exact quasinormal mode spectrum and eigenfunctions in the large black hole limit. From this, we get several new and surprising results.
\begin{enumerate} 
\item The spectrum is fully determined by imposing causality at the black hole boundary. The boundary conditions only determine the form of the eigenfunctions.
\item The spectrum, properly rescaled, turns out to be of the form
\begin{equation}
0, - \iu, - 2 \iu, \dots, - n \iu, \dots, \qquad n \in \mathbb{Z},
\label{eq:spectrumLBL}
\end{equation}
where the existence of the zero mode depends on whether the boundary conditions at spatial infinity can support a nonzero constant solution eigenfunction. 
\item An expansion of the spectrum around the large black hole limit may be derived agnostic of a boundary condition at spatial infinity. This allows a proper quantitative characterization of how insensitive the spectrum is to the boundary condition at infinity, in the large black hole limit.
\end{enumerate}

As a consequence of \eqref{eq:spectrumLBL}, electromagnetic perturbations cannot vibrate in the large black hole limit. When we reverse the rescaling, the spectrum corresponds to infinitely damped modes, which seems to play an important role in theories for quantum gravity \cite{hod1998, dreyer2004, cardoso2004}.

Third, we shall show that not only is the spectrum strongly independent of $\ell$ in the large black hole limit, but that it is prescribed by a single number
\begin{equation}
\gamma = \ell(\ell+1)\rho^2.
\end{equation}
These observations shall reveal a new symmetry of the spectrum in the large black hole limit: \emph{spectral similarity}. This property is distinct from isospectrality since the spectra calculated for different pairs of $(\rho, \ell)$ are equal to within an order $\textit{O}(\rho^{2}_{\max})$, where $\rho_{\max}$ is the largest value of $\rho$ in that set of pairs.

Our results are both analytic and numerical. Where the numerical results are concerned, a novel pseudospectral method was used by using the Bernstein polynomial basis. The code we used is distributed as a \texttt{Mathematica} package we call \texttt{SpectralBP} \cite{fortuna2021}.

This paper is organized as follows. In Sec. \ref{sec:eqandnum}, we introduce the Regge-Wheeler master equation, and manipulate it for numerical and analytical use. We will also briefly describe the pseudospectral method we implemented. In Sec. \ref{sec:numres}, we numerically recreate well-known results in the literature, as well as add a few observations and comments relating to the bifurcations.

In Sec. \ref{sec:selfsim}, we provide numerical evidence of the spectral similarity of the spectrum in the large black hole limit. We explain this in terms of a perturbed eigenvalue problem we may reach in the same limit. In Sec. \ref{sec:master}, we use this spectral similarity and some numerics to show the existence of the Feigenbaum-like constant determining the bifurcations and derive a master equation for when the bifurcations occur.

Finally, in Sec. \ref{sec:LBHL}, we derive exact solutions in the large black hole limit and the expansion of the spectrum around the same limit. We also show the insensitivity of the spectrum to the boundary conditions at spatial infinity.

\section{Equations and numerical method}
\label{sec:eqandnum}

In this section, we shall derive the ordinary differential equation (ODE) eigenvalue problem used in the analysis of the proceeding sections. We shall be using natural units $(G = c = 1)$, and denoting dimensionful quantities with barred variables and dimensionless quantities with unbarred variables.

When a spherically symmetric spacetime in Schwarzschild coordinates,
\begin{equation}
ds^2 = -f(\bar{r}) d\bar{t}^2 + f(\bar{r})^{-1} d\bar{r}^2 + \bar{r}^2 d\Omega^2,
\label{eq:GenSpherST}
\end{equation}
interacts linearly with an external field, its perturbations may be described by the Regge-Wheeler equation,
\begin{equation}
\partial_{\bar{t}}^2 \Phi_{s,\ell} + (- \partial_{\bar{x}}^2 + V_{s,\ell}) \Phi_{s,\ell} = 0,
\label{eq:ReggeWheelertx}
\end{equation}
where $s$ is the spin weight of the perturbing field, $\ell \geq s$ is a harmonic mode index, $\bar{x}$ is the Regge-Wheeler tortoise of the spacetime \eqref{eq:GenSpherST} and $V_{s,\ell}$ is an effective potential dependent on the spin weight of the external field and the metric function $f(\bar{r})$.

Solutions with a stationary wave ansatz,
\begin{equation}
\Phi_{s,\ell} = R_{\bar{\omega}, s, \ell}(\bar{x}) \exp(-i \bar{\omega} t),
\end{equation}
together with appropriate boundary conditions that impose physical requirements such as causality or energy conservation are called quasinormal modes. These modes form a discrete and generally complex spectrum, where the nonzero imaginary component of the spectrum signifies that perturbations may decay either through black hole boundary or out to spatial infinity.

In tortoise coordinates, quasinormal modes are solutions to a Schr\"{o}dinger-like equation,
\begin{equation}
\left( - \deri{^2}{\bar{x}^2} + V_{s,\ell} \right) R_{\bar{\omega}, s, \ell} = \bar{\omega}^2 R_{\bar{\omega}, s, \ell}.
\label{eq:ReggeWheelerx}
\end{equation}
For an SAdS black hole, the metric function has the form
\begin{equation}
f(\bar{r}) = 1 - \dfrac{2\bar{M}_h}{\bar{r}} + \dfrac{\bar{r}^2}{\bar{L}^2},
\end{equation}
where $\bar{M}_h$ is the mass of the Schwarzschild black hole and $\bar{L}$ is the AdS curvature radius.

For analysis, the location of the black hole horizon $\bar{r}_h$ is more important than knowing its mass. Moving forward, we replace
\begin{equation}
\bar{M}_h \to \dfrac{\bar{r}_h(\bar{L}^2 + \bar{r}_h^2)}{2\bar{L}^2},
\end{equation}
so that
\begin{equation}
f(\bar{r}) = 1 - \left( 1 + \dfrac{\bar{r}_h^2}{\bar{L}^2} \right) \dfrac{\bar{r}_h}{\bar{r}} + \dfrac{\bar{r}^2}{\bar{L}^2}.
\label{eq:SAdSmetricfunc}
\end{equation}

\subsection{Boundary conditions}

For asymptotically flat spacetimes, the effective potential $V_{s,\ell}$ vanishes for any given $s$ and $\ell$ as you approach either horizon,
\begin{equation}
\lim_{\bar{x} \to \pm \infty} V_{s,\ell}(\bar{x}) = 0.
\end{equation}
Thus, it is reasonable for any perturbation with compact support outside the black hole horizon to impose the causal requirement that perturbations may only fall into the black hole or radiate out to infinity,
\begin{equation}
\begin{matrix*}[l]
\Phi_{s,\ell}(\bar{x}, \bar{t}) \sim \exp( -\iu \bar{\omega} (\bar{t} + \bar{x}) ), & \bar{x} \to -\infty, \\
\Phi_{s,\ell}(\bar{x}, \bar{t}) \sim \exp( -\iu \bar{\omega} (\bar{t} - \bar{x}) ), & \bar{x} \to +\infty. \\
\end{matrix*}
\end{equation}

For anti--de Sitter spacetimes, however, we lose one of these physically motivated constraints. This is because of the cosmological term in the metric function,
\begin{equation}
\lim_{\bar{r} \to \infty} f(\bar{r}) \sim \dfrac{\bar{r}^2}{\bar{L}^2}.
\end{equation}
The effective potential for an SAdS spacetime may either diverge (scalar or gravitational perturbations) or become a positive nonzero value (electromagnetic perturbations)  as $\bar{x} \to \infty$.

Thus, perturbations with compact support should be a mixture of ingoing and outgoing plane waves as you approach spatial infinity even accounting for causality. Since our Universe is not asymptotically AdS, there is no obvious physics that motivates a particular boundary condition at spatial infinity. We have a free choice, usually motivated by numerical or analytical convenience or by considerations coming from the AdS/CFT duality \cite{son2002,michalogiorgakis2007,skenderis2009,bakas2009,bakas2009a}.

Here, we choose to impose that the ingoing and outgoing waves at spatial infinity exactly cancel out, as in a  total internal reflection.
\begin{equation}
\begin{matrix*}[l]
\Phi_{s,\ell}(\bar{x}, \bar{t}) \sim \exp( -\iu \bar{\omega} (\bar{t} + \bar{x}) ), & \bar{x} \to -\infty, \\
\Phi_{s,\ell}(\bar{x}, \bar{t}) \sim 0, & \bar{x} \to +\infty. \\
\end{matrix*}
\label{eq:bccausal}
\end{equation}
We make this choice, at first, for convenience. The literature seems to favor Robin boundary conditions at spatial infinity for AdS spacetimes.  However, as we shall see, the spectrum of the EM-SAdS black hole in the large black hole limit is insensitive to the choice of boundary conditions at spatial infinity.

\subsection{The eigenvalue problem}
For the calculations that will follow, it is more convenient to solve Eq.~\eqref{eq:ReggeWheelerx} using the Schwarzschild $\bar{r}$, defined implicitly by
\begin{equation}
\deri{\bar{x}}{\bar{r}} = f(\bar{r})^{-1},
\end{equation}
since the potential $V_{s,\ell}(\bar{x})$ has a complicated form in tortoise coordinates. For EM SAdS, we choose $s = 1$, which has the form
\begin{equation}
V_{1,\ell}(\bar{r}) = f(\bar{r}) \dfrac{\ell(\ell+1)}{\bar{r}^2}
\label{eq:EMPotential}
\end{equation}
and $f(\bar{r})$ is the metric Schwarzschild-AdS metric function given in \eqref{eq:SAdSmetricfunc}.

The black hole radius $\bar{r}_h$ and the AdS radius $\bar{L}$ provide natural length scales to nondimensionalize our variables. In this paper, we shall be freely switching between two coordinates: (1) a dimensionless coordinate system where $r = 1$ corresponds to the black hole horizon,
\begin{equation}
\bar{r} \to \bar{r}_h r, \qquad \bar{L} \to \bar{r}_h L, \qquad \bar{\omega} \to \dfrac{\omega}{\bar{r}_h},
\label{eq:dimensionless1}
\end{equation}
and (2) a dimensionless coordinate system normalized to the AdS radius,
\begin{equation}
\bar{r} \to \bar{L} r, \qquad \tilde{r}_h \to \bar{L} r_h, \qquad \bar{\omega} \to \dfrac{\omega}{\bar{L}}.
\label{eq:dimensionless2}
\end{equation}
In either dimensionless coordinate system, the dimensionless quantity
\begin{equation}
\rho = \dfrac{\bar{L}}{\bar{r}_h}
\end{equation}
is constant. We shall use this dimensionless quantity to speak about both coordinate systems simultaneously, where it may mean as the dimensionless AdS radius in coordinates described by \eqref{eq:dimensionless1} or as the inverse dimensionless black hole radius in coordinates described by \eqref{eq:dimensionless2},
\begin{equation}
\eqref{eq:dimensionless1} \to \rho = L, \qquad \eqref{eq:dimensionless2} \to \rho = r_h^{-1}.
\label{eq:dimensionlessconversion}
\end{equation}
We note that the large black hole limit then corresponds to
\begin{equation}
\rho \to 0.
\end{equation}
In either dimensionless coordinate system, the domain of the solution is semi-infinite. In particular, the semi-infinite regions correspond to $(1,\infty)$ and $(\rho^{-1},\infty)$ when normalizing to the black hole radius and the AdS radius, respectively.

We choose a transformation so that both semi-infinite intervals are mapped to the same compact interval $(0,1)$. We use the coordinate transformation
\begin{equation}
r = \dfrac{1}{u}
\end{equation}
for the coordinates normalized to the black hole radius, and
\begin{equation}
r = \dfrac{1}{\rho u}
\end{equation}
for the coordinates normalized to the AdS radius.
These manipulations starting from Eq.~\eqref{eq:ReggeWheelertx} yield the same differential equation 
\begin{multline}
u^4 f(u)^2 R_{\omega,\ell}''(u) + u^3f(u)\left(2 f(u) + u f'(u)\right) R_{\omega,\ell}'(u) + \\
 \left( \omega^2 - \ell(\ell+1)f(u) \right) R_{\omega,\ell}(u) = 0
\end{multline}
We proceed to peel away the singular behavior at the black hole boundary
\begin{equation}
R_{\omega, \ell}(u) = \exp(- i \omega x(u)) \phi_{\omega,\ell}(u)
\label{eq:rescaling}
\end{equation}
The equation now becomes
\begin{dmath}
u^2 f(u) \deri{^2 \phi_{\omega,\ell}(u)}{u^2} + \left( 2 \iu \omega + 2 u f(u)
+ u^2 \deri{f(u)}{u} \right) \deri{\phi_{\omega,\ell}(u)}{u} 
- \ell(\ell+1)\phi_{\omega,\ell}(u) = 0.
\label{eq:ReggeWheeleru}
\end{dmath}
As a sanity check, consider a Frobenius expansion around the black hole horizon $u = 1$ with
\begin{equation}
\phi_{\omega,\ell}(u) = (u-1)^y \sum_{n=0}^\infty c_n (u-1)^n,
\end{equation}
where $y$ is the indicial exponent, solved by the indicial equation
\begin{equation}
y(y-1) + ay + b = 0
\end{equation}
and
\begin{equation}
\begin{aligned}
a &= \lim_{u\to1} (u-1) \dfrac{2 \iu \omega + 2 u f(u) + u^2 \deri{f(u)}{u}}{u^2 f(u)} \\ 
b &= \lim_{u \to 1} (u-1)^2 \dfrac{- \ell(\ell+1)}{u^2 f(u)}.
\end{aligned}
\end{equation}
That is,
\begin{equation}
a = \dfrac{3+ \rho^2 - 2 \iu \omega \rho^2}{3 + \rho^2}, \qquad b = 0,
\end{equation}
The indicial equation has two solutions,
\begin{equation}
y = 0, \qquad \dfrac{2 \iu \omega \rho^2}{3 + \rho^2}.
\end{equation}
This means there are two solutions around the black hole horizon, with the asymptotic behavior,
\begin{equation}
\begin{matrix*}[l]
\phi^{+}_{\omega,\ell}(u) & \sim (u-1)^{+ \iu \omega \frac{2 \rho^2 }{3 + \rho^2}}  \\
\phi^{-}_{\omega,\ell}(u) & \sim 1
\end{matrix*}
\end{equation}
We note that we have successfully scaled out the causal part, since the first solution may be seen as exiting the black hole horizon when we bring back the time dependence.

Curiously, the surface gravity \cite{horowitz2000}
\begin{equation}
\kappa = 2 \pi T,
\end{equation}
where $T$ is the dimensionless Hawking temperature given by
\begin{equation}
T = \dfrac{3 + \rho^2}{4 \pi \rho^2},
\end{equation}
appears in the indicial exponent of the acausal solution,
\begin{equation}
y = 0, \qquad \dfrac{\iu \omega}{\kappa}.
\end{equation}
This motivates us to rescale $\omega$ with
\begin{equation}
\omega \to \left(\dfrac{3+\rho^2}{2\rho^2}\right)\lambda.
\end{equation}
This scales out the diverging behavior of $\omega$ in the large black hole limit, leaving $\lambda$ finite for all values of $\rho$. This rescaled frequency, $\lambda$, will feature heavily in the rest of the paper.

The differential equation finally becomes
\begin{multline}
(u-1)\left( (1 + \rho^2) u^2 + u + 1 \right) \deri{^2 \phi_{\lambda,\ell}(u)}{u^2} \\
+ \left( - \iu (3 + \rho^2) \lambda - 2 \rho^2 u + 3 (1 + \rho^2) u^2 \right) \deri{\phi_{\lambda,\ell}(u)}{u} \\
+ \ell(\ell+1)\rho^2 \phi_{\lambda,\ell}(u) = 0,
\label{eq:ReggeWheelerEMSAdS}
\end{multline}
where the domain of the solution and the boundary conditions are identical for either coordinate system.

To illustrate the usefulness of \eqref{eq:ReggeWheelerEMSAdS}, consider, for example, Fig. \ref{fig:bifurcation2} below. There are two main features we needed to resolve more accurately so that the interpolated graph looked smooth. These are the first turning points of the real part of the QNM frequencies, and the second are the bifurcation points where the graph transitions from a locally linear function to two branches of a square root function, or vice versa -- depending on whether you are looking at the imaginary or real part.

The first feature is located where $\rho^{-1}$ is small and the second feature is located where $\rho^{-1}$ is large. Thus, it was convenient for us to search for points with coordinates normalized to $\bar{r}_h$ for the first feature and coordinates normalized to $\bar{L}$ for the second feature. We then used the fact that $\rho$ is constant in either coordinate system to construct Fig. \ref{fig:bifurcation2}.

\subsection{Numerical method}
\label{subsec:nummeth}

Hereafter, we shall solve several ODE eigenvalue problems using a collocation method, expanding the eigenfunction $\phi_{\lambda,l}(u)$ as a linear sum of weighted basis functions,
\begin{equation}
\phi_{\lambda,\ell}(u) = \sum_{k=0}^N c_k B_k^N(u).
\label{eq:basis}
\end{equation}
Specifically, we have used the Bernstein polynomial basis, given by
\begin{equation}
B_k^N(u) = \begin{pmatrix}
N \\ k
\end{pmatrix} u^k(1-u)^{N-k}, \,\, \begin{pmatrix}
N \\ k
\end{pmatrix} = \dfrac{N!}{k!(N-k)!}.
\end{equation}

When the expansion \eqref{eq:basis} is plugged into \eqref{eq:ReggeWheelerEMSAdS} and evaluated at a set of collocation points, this turns the ODE eigenvalue problem \eqref{eq:ReggeWheelerEMSAdS} together with the total internal reflection boundary condition,
\begin{equation}
\phi_{\lambda,\ell}(u) \sim u, \qquad u \to 0,
\label{eq:TIRBC}
\end{equation}
into a generalized eigenvalue problem for the set of coefficients $c_k$,
\begin{equation}
\textbf{M}(\lambda) \textbf{c} = 0.
\label{eq:matrixequation}
\end{equation}
Significant digits for $\lambda$ are determined by calculating some particular value using two values of $N$ in \eqref{eq:basis} sufficiently far apart, keeping digits that are common between the two calculations. For more details, please refer to \cite{fortuna2021}.

The code we have used is implemented in a \texttt{Mathematica} package we call \texttt{SpectralBP}. This Bernstein spectral implementation and all its unappreciated advantages are fully described in \cite{fortuna2021}. A summary of its advantages and disadvantages is feature in Sec. V therein. 

As an example, the Bernstein basis has a special property where a general class of mixed boundary conditions may decouple a set of coefficients $c_k$, and may be solved independently of the differential equation. This leads to many computational conveniences (enumerated in \cite{fortuna2021}), such as the exact satisfaction of boundary conditions such as \eqref{eq:TIRBC} and the reduction of the size of the matrix equation \eqref{eq:matrixequation}.

In contrast, the boundary conditions do not generally decouple a set of coefficients for other basis polynomials such as Chebyschev polynomials, and the algebraic equations which the boundary conditions impose on the set of coefficients must be solved with the differential equation as in a tau method. Because of the limitations of floating point arithmetic, these boundary conditions will not be generally exactly satisfied as well, unless special measures are implemented.

\section{Bifurcations of the spectrum}
\label{sec:numres}

\begin{figure}[t!]
\begin{center}
\includegraphics[width=0.44\textwidth]{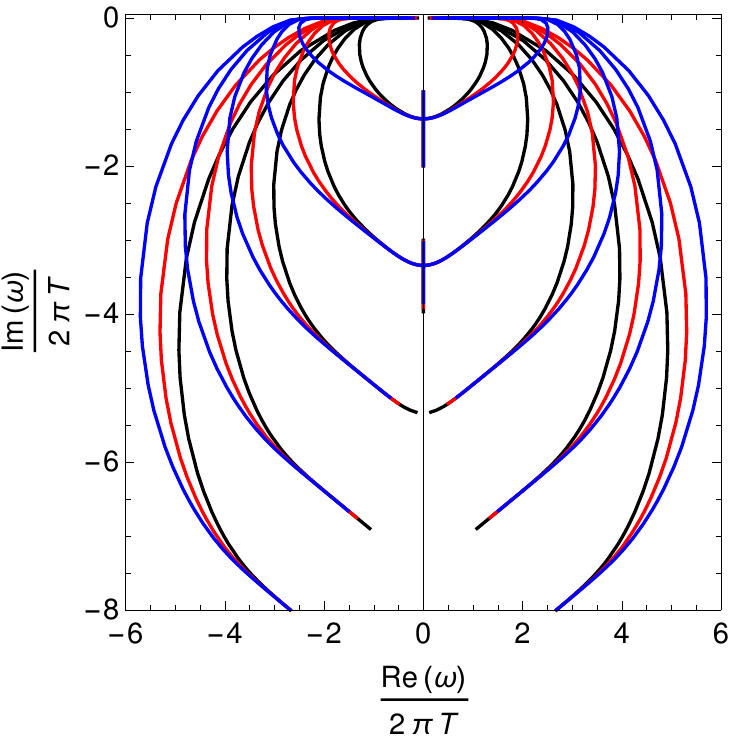}
\end{center}
\caption{EM-SAdS quasinormal modes plotted on the complex plane, for $\ell=1$ (black lines) $\ell = 2$ (red lines) and $\ell=3$ (blue lines) and $\rho \in \left(10^{-2},50\right)$. Different curves correspond to different overtones.}
\label{fig:freqs}
\end{figure}

Figure \ref{fig:freqs} shows the resulting eigenvalues plotted on the complex plane for different values of $\ell$, varying the value of $\rho \in\left(10^{-2}, 50\right)$. The spectra calculated here match prior numerical work \cite{cardoso2001, cardoso2003} when the proper units and scaling are folded back in. \footnote{Table I and II of \cite{cardoso2001} and Table V-VIII of \cite{cardoso2003}} We also confirm the following observations from the recent literature \cite{wang2021}: When a pair of eigenvalues satisfying the symmetry \eqref{eq:symmetry} meet at the negative imaginary axis, they split into two purely imaginary or overdamped modes.

Such overdamped modes have been observed before in other spacetimes, such as the Kerr metric \cite{cook2014,cook2016}. As the spin parameter of the Kerr metric is increased, quasinormal mode frequencies are seen to fall into and emerge out of the negative imaginary axis. 

There is a related splitting that occurs where a suite (not just a pair) of quasinormal mode frequencies limit to a set of equally spaced modes on the real axis. This was first analytically predicted to occur near the extremal Kerr limit \cite{hod2008a,hod2008,hod2009}, where the limit set is dependent only on the azimuthal harmonic index $m$ and the Kerr black hole's mass. This was later numerically confirmed in high accuracy numerical studies \cite{yang2013a, yang2013, cook2014}. 

Both scenarios are distinct from the quasinormal mode splitting that occurs around the Schwarzschild spectrum when the black hole is given spin.

Explicit observation of what these overdamped modes are doing on the negative imaginary axis seems to have been hindered by methods that cannot find purely imaginary eigenvalues \cite{cook2016}. However, in the Schwarzschild--de Sitter spacetime, the behavior of these overdamped modes have been observed using a purely spectral code \cite{jansen2017}. 

As the ratio between the de Sitter and black hole radii is increased, a pair of QNM modes satisfying \eqref{eq:symmetry} meet up at the negative imaginary axis. There, these overdamped modes stay, moving up and down the negative imaginary axis until they collide with another overdamped mode. The reverse of the bifurcation then occurs: the two overdamped modes leave the negative imaginary axis as two pairs satisfying \eqref{eq:symmetry}. We refer to Fig. 8 of Ref. \cite{jansen2017} for an example of this.

We point out three new observations pertinent to our study:
\begin{enumerate}
\item The rich behavior of the quasinormal modes separating and recolliding at the negative imaginary axis is not general. In contrast to the Kerr and de Sitter cases mode splitting, here we find a scenario where the overdamped modes approach a limit set on the negative imaginary axis.
\item These bifurcations occur at approximately the same point between $[-(2n+1) \iu, - (2n+2) \iu]$, independent of $\ell$.
\item One of the eigenvalues approaches $-(2n+1)\iu$ and the other approach $-(2n+2) \iu$ in the large black hole limit. In other words, entire spectrum  approaches
\begin{equation}
\lambda \to - \iu, - 2\iu, -3 \iu, \dots, \qquad \rho \to 0.
\end{equation}
\end{enumerate}

In later sections, we shall provide analytic foundations to these numerical observations.

\section{Spectral similarity and $\ell$ independence}
\label{sec:selfsim}

We first demonstrate that the spectrum exhibits a property in the large black hole limit which we call spectral similarity. To the best of our knowledge, this is a newly discovered symmetry of the spectrum which clarifies that the $\ell$ independence and the equal spacing of the spectrum are independent properties.

Let $\{\lambda_{n,l,\rho}\}$ be the spectrum of \eqref{eq:ReggeWheelerEMSAdS} for some $\ell$ and $\rho$, and let
\begin{equation}
\gamma(\ell, \rho) = \ell(\ell+1) \rho^2.
\label{eq:selfsimilarity2}
\end{equation}
We find that for any pairs of $(\ell, \rho)$ and $(\ell', \rho')$ satisfying
\begin{equation}
\gamma(\ell, \rho) = \gamma(\ell', \rho'), \qquad \rho \ll 1,
\end{equation}
the two respective spectra are approximately equal,
\begin{equation}
\lambda_{n,\ell, \rho} \approx \lambda_{n, \ell', \rho'}.
\end{equation}

\begin{figure}[t]
\begin{centering}
\begin{tabular}{c}
  \includegraphics[width=0.4\textwidth]{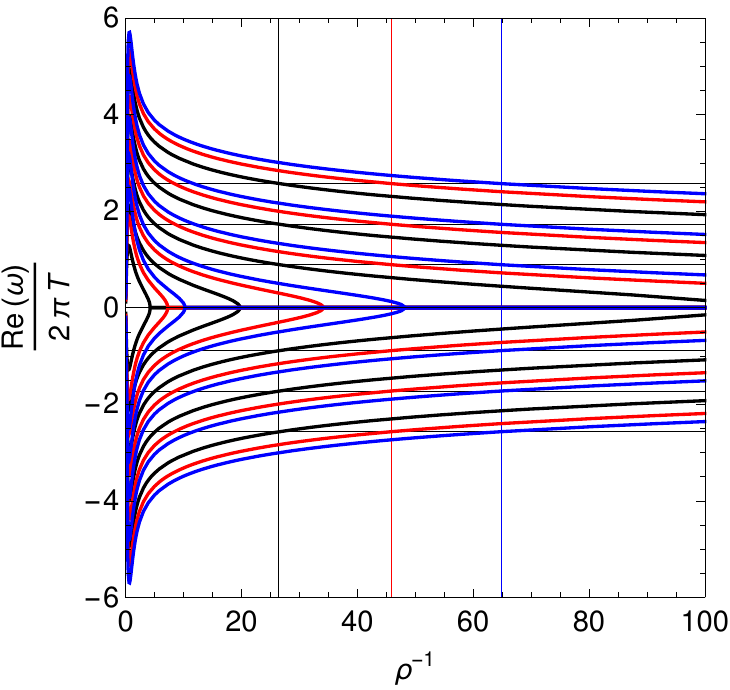} \\  \includegraphics[width=0.4\textwidth]{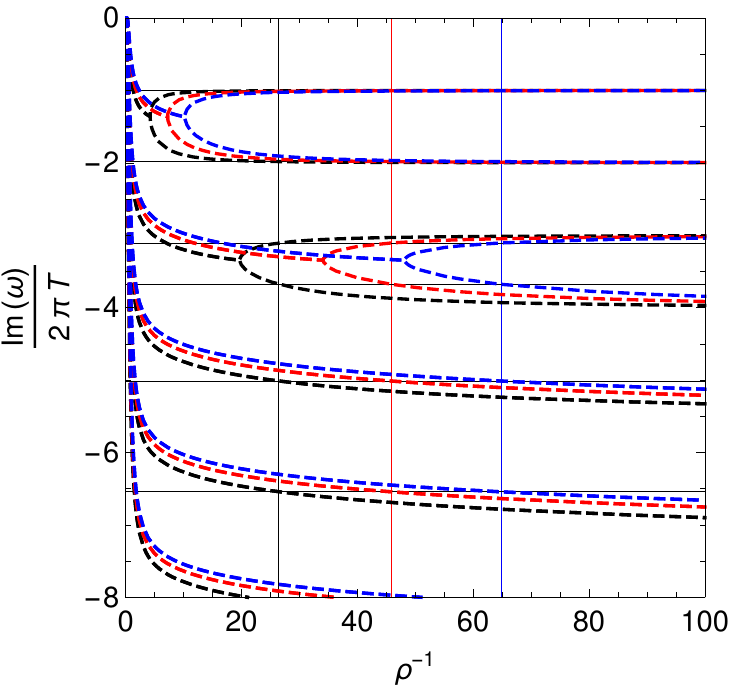}
\end{tabular}
\caption{Real and imaginary parts of the eigenvalues as a function of $\rho^{-1}$, with the same color scheme as Fig. \ref{fig:freqs}. Vertical black, red and blue lines are $\rho$ that satisfy $\gamma = 1/350 =\ell(\ell+1)\rho^2$ for $\ell = 1, 2, 3$ respectively. Horizontal lines serve as guides to show that the spectra are similar at the different values of $\rho$ and $l$.}
\label{fig:bifurcation2}
\end{centering}
\end{figure}

\begin{figure}[t]
\begin{centering}
\begin{tabular}{c}
  \includegraphics[width=0.4\textwidth]{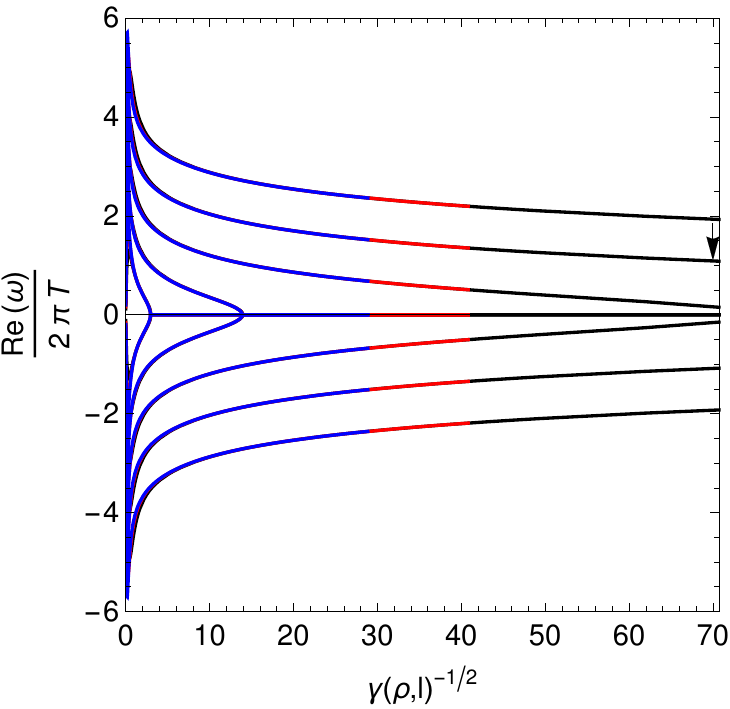} \\  \includegraphics[width=0.4\textwidth]{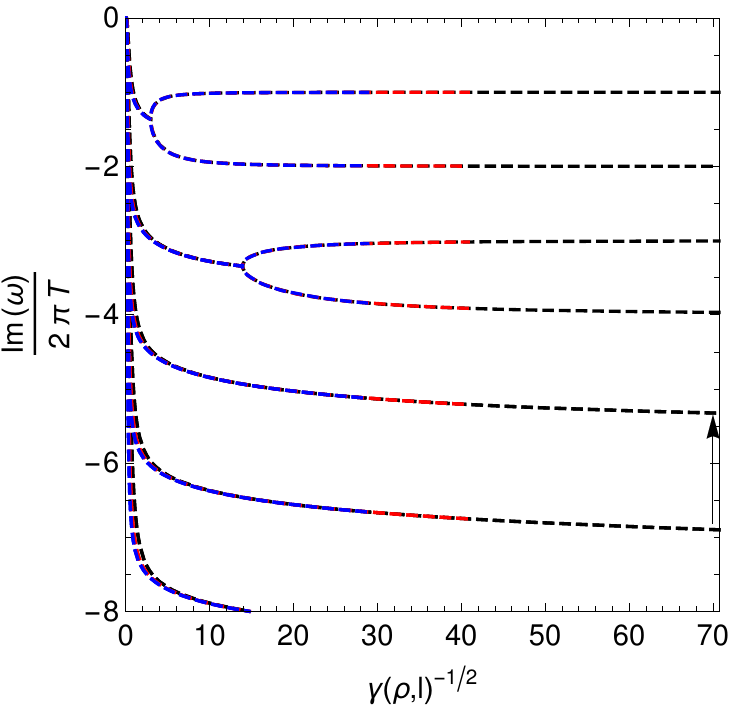}
\end{tabular}
\caption{Real and imaginary parts of the eigenvalues as a function of the parameter $\gamma(\rho,\ell)^{-1/2}$, with the same color scheme and data as Fig. \ref{fig:bifurcation2}. We note that the same interval in Fig. \ref{fig:bifurcation2} is mapped to different intervals in $\gamma(\rho,\ell)^{-1/2}$ for different values of $\ell$. Specifically, $\rho \in (10^{-2},50)$ is mapped to (0.01414, 70.71) for $\ell = 1$, to (0.008165, 40.8248) for $\ell = 2$ and to (0.005774, 28.87) for $\ell = 3$. The coincidence of the three curves in the large black hole limit is predicted by \eqref{eq:l-independence}, where we note that $\gamma(\rho,\ell)^{-1/2} \sim \rho^{-1}$ in the same limit. This indicates the spectral similarity of the spectrum. The gap indicated by the arrows is $0.845 - 1.52 \iu$.}
\label{fig:bifurcation3}
\end{centering}
\end{figure}

One already sees hints of spectral similarity in Fig.~\ref{fig:bifurcation2}, but it is most strongly demonstrated in Fig.~\ref{fig:bifurcation3}. In Fig.~\ref{fig:bifurcation2}, we show the dependence of the spectrum to both $\rho$ and three different values of $\ell$. We have included guiding lines for a specific example of spectral similarity, choosing $\gamma(\ell, \rho) = 1/350$ and marking with vertical lines where $\rho^{-1}$ satisfies the prior expression when $\ell = 1, 2, 3$. The intersection of a vertical line of a color to their respective colored curve represents the spectrum for that particular $\ell$. All three spectra are approximately equal, indicated by the horizontal lines where at least visually all three spectra `hit'.

We note that this approximate equality extends to all overtone numbers, not just the eigenvalues we have plotted in Fig.~\ref{fig:bifurcation2}. We also note that this approximate equality shows up for all values of $\gamma(\ell,\rho)$ that are sufficiently small enough.

This spectral similarity is more distinctly shown in Fig.~\ref{fig:bifurcation3}, where we parametrized the spectra for different values of $\rho$ and $\ell$ using \eqref{eq:selfsimilarity2}. Spectral similarity manifests when different values of $\rho$ and $\ell$ almost coincide whenever the corresponding values of $\gamma(\rho,\ell)$ are equal.

To compare with Fig.~\ref{fig:bifurcation2}, we use $\gamma(\rho,\ell)^{-1/2}$ since
\begin{equation}
\gamma(\rho,\ell)^{-1/2} \sim \rho^{-1}.
\end{equation}

The coincidence of the three curves corresponding to $\ell = 1, 2, 3$ in the large black hole limit is predicted by \eqref{eq:l-independence}, where we note that $\gamma(\rho,\ell)^{-1/2} \sim \rho^{-1}$ in the same limit. We emphasize that the coincidence is not exact, which is why we refrain from using the term isospectral for this property. As we shall see, for different values of $\ell$, the difference between similar spectra is bounded by the square of the largest value of $\rho$ in that set.

We note that the same interval in $\rho$ is mapped to different intervals in $\gamma(\rho,\ell)^{-1/2}$ for different values of $\ell$. Specifically, $\rho \in (10^{-2},50)$ is mapped to (0.01414, 70.71) for $\ell = 1$, to (0.008165, 40.8248) for $\ell = 2$ and to (0.005774, 28.87) for $\ell = 3$.

One may also observe that for small and intermediate black holes, this symmetry is broken. Note how the spectra split into different curves in Fig.~\ref{fig:bifurcation3} for $\gamma(\rho,\ell)^{-1/2} \leq 5$.

To explain spectral similarity, we may define a related eigenvalue problem to  \eqref{eq:ReggeWheelerEMSAdS} by simultaneously reaching the eikonal limit $\ell \to \infty$ and the large black hole limit $\rho \to 0$, while keeping the product
\begin{equation}
\gamma = \ell(\ell+1)\rho^2 
\label{eq:EMSAdSgamma}
\end{equation}
constant. We would arrive at
\begin{multline}
\left( u^3 - 1 \right) \deri{^2 \phi_{\lambda,\gamma}(u)}{u^2} + 3 \left( u^2 - \iu \lambda \right) \deri{\phi_{\lambda,\gamma}(u)}{u} \\
+ \gamma \phi_{\lambda,\gamma}(u) = 0.
\label{eq:ReggeWheelerEMSAdSELBL}
\end{multline}
Then $\{\lambda_{n,\ell,\rho}\}$ of \eqref{eq:ReggeWheelerEMSAdS} is approximately equal to $\{\lambda_{n,\gamma}\}$ of \eqref{eq:ReggeWheelerEMSAdSELBL} in the sense that
\begin{equation}
\left| \lambda_{n,\gamma} - \lambda_{n,\ell,\rho} \right| \sim \mathcal{O}(\rho^{2})
\label{eq:l-independence}
\end{equation}
or
\begin{equation}
\left| \lambda_{n,\gamma} - \lambda_{n,\ell,\rho} \right| \sim \mathcal{O}\left(\dfrac{\gamma}{l(l+1)}\right)
\label{eq:l-independence2}
\end{equation}
This is because \eqref{eq:ReggeWheelerEMSAdS} may be treated as a perturbed eigenvalue problem of \eqref{eq:ReggeWheelerEMSAdSELBL}, as in
\begin{equation}
\hat{\mathfrak{L}}_1(u,\gamma,\lambda)\phi(u) + \delta^2 \hat{\mathfrak{L}}_2(u,\lambda) \phi(u) = 0,
\label{eq:perturbed}
\end{equation}
where $\hat{\mathfrak{L}}_1(u,\gamma,\lambda)\phi(u)$ is given by \eqref{eq:ReggeWheelerEMSAdSELBL} with perturbing parameter $\delta = \rho$ and
\begin{multline}
\hat{\mathfrak{L}}_2(u,\lambda) \phi(u) = u^2(u-1) \deri{^2 \phi(u)}{u^2} \\
+ \left( - \iu \lambda - 2 u + 3 u^2 \right) \deri{\phi_{\lambda,l}(u)}{u}.
\end{multline}
The perturbation embodies finite-size effects of the black hole to the eigenvalue problem. That is, we may expand
\begin{equation}
\phi(u) = \psi_0(u) + \sum_{k=1}^\infty \delta^{2k} \psi_k(u), \; \lambda = \mu_0 + \sum_{k=1}^\infty \delta^{2k} \mu_k,
\end{equation}
where $(\psi_0(u),\mu_0)$ are the eigensolutions of the unperturbed ODE eigenvalue problem
\begin{equation}
\hat{\mathfrak{L}}_1(u,\gamma,\mu_0)\psi_0(u) = 0.
\label{eq:unperturbed}
\end{equation}

Spectral similarity is sufficient in explaining the strong $\ell$ independence of the spectrum: in the large black hole limit, Eq. \eqref{eq:l-independence} tells us that it is $\gamma$ which determines the form of the spectrum with corrections scaling with $\sim \rho^2$. In fact, Eq. \eqref{eq:l-independence2} tells us that the larger the value of $\ell$, the quicker the independence sets it (for the same value of $\gamma$).

On the other hand, the equal spacing of the spectrum is orthogonal to $\ell$ independence, contingent on the properties of \eqref{eq:ReggeWheelerEMSAdSELBL}. One can imagine a different form of $\mathfrak{L}_1(u,\gamma,\lambda)$ which results with a spectrum that is unequally spaced. This would result in a spectrum of the original problem being both $\ell$ independent and unequally spaced.

In the ``asymptotic limit" \cite{cardoso2004} (i.e., $n \to \infty$ and $\rho \to \infty$), the spacing for general asymptotically AdS spacetimes has been shown to be 
\begin{equation}
\lambda_{n+1,\ell,\rho} - \lambda_{n,\ell,\rho} = \dfrac{\sqrt{3}}{2} - \dfrac{3 \iu}{2} = 0.866025 - 1.5 \iu.
\label{eq:asymptoticspacing}
\end{equation}
We have already applied our specific scaling for better comparison. This equal spacing is visible in Fig. \ref{fig:bifurcation3}, where even the low overtones match the asymptotic value reasonably well.

The spacing of \eqref{eq:ReggeWheelerEMSAdSELBL} limits to \eqref{eq:asymptoticspacing}. For example, choosing $\gamma = 10^{-24}$, we have
\begin{equation}
\lambda_{71,\gamma} - \lambda_{70,\gamma} = 0.864(2560721) - 1.5014(08288) \iu. \footnote{Significant digits, as stated in Sec. \ref{subsec:nummeth}, is determined by calculating two spectra using two values of $N$ in \eqref{eq:basis}. Specifically, we have used N = 550 and N = 600. We have also included in brackets digits coming from the higher accuracy calculation.}
\end{equation}

\subsection{Comments on `bifurcation'}
\label{subsec:bifurcation}

\begin{figure}[b!]
\begin{center}
\includegraphics[width=0.44\textwidth]{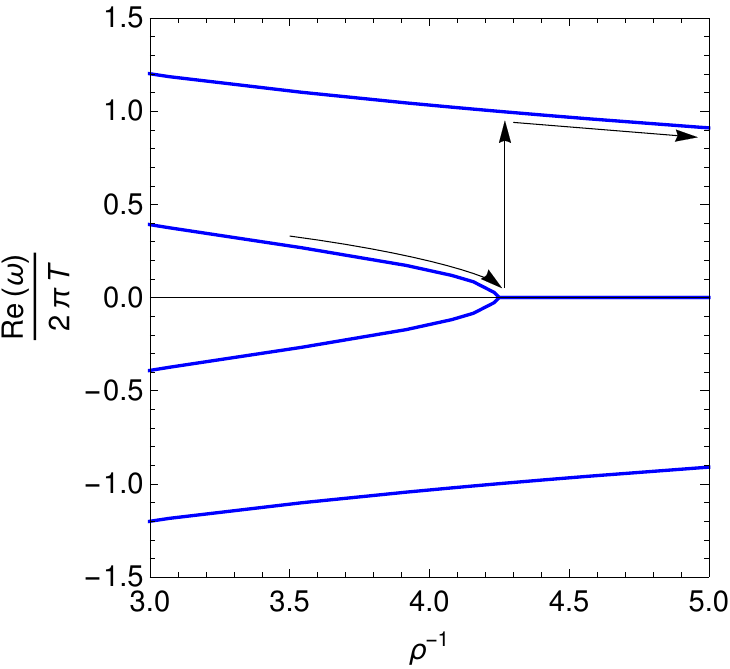}
\end{center}
\caption{Bifurcation scenario where the appearance of an overdamped mode implies a jump in the frequency of the lowest lying vibrational mode. As the parameter is varied around $\rho^{-1} \sim 4.3$, the period of the lowest lying vibrational mode may either diverge or remain finite, depending on which side of the bifurcation you are on.}
\label{fig:helpfulfigure}
\end{figure}

The description of the splitting of two pairs satisfying \eqref{eq:symmetry} into two purely imaginary modes as ``bifurcations" was given very recently \cite{wang2021, wang2021a}. (In the earlier Ref.~\cite{jansen2017}, this mode splitting was not referred to as a bifurcation.) A bifurcation implies that a small continuous change in a parameter of a system results in a sudden change in the qualities of that same system, so it matters to specify what discontinuously changes as $\rho$ is varied. 

It is not difficult to justify why the description of \cite{wang2021, wang2021a} is appropriate.

When two quasinormal modes satisfy the symmetry \eqref{eq:symmetry}, the split into two distinct purely imaginary modes breaks the symmetry. Before and after the bifurcation, the number of pairs that satisfies \eqref{eq:symmetry} changes. 

This should also shift the overtone number labeling of all modes above them, and their corresponding eigenfunctions no longer satisfy the symmetry \eqref{eq:symmetry} as well.

Furthermore, if you imagine how electromagnetic perturbations vibrate outside the black hole, the fundamental ``note" of these perturbations jumps before and after the bifurcation. Specifically, the period of the lowest lying vibration mode diverges prior to the bifurcation, while the period of the lowest lying vibration mode is finite after the bifurcation.

This is illustrated more clearly in Fig. \ref{fig:helpfulfigure}, which shows the path and jump of the fundamental vibrational frequency around a bifurcation for $\ell = 1$.

Finally, as demonstrated in the prior discussion, there are actually two spacings to which the spectrum limits: the well-known one describes the spacing of the unbifurcated modes. The second spacing describes the bifurcated modes, given by
\begin{equation}
\lambda_{n+1,\gamma} - \lambda_{n,\gamma} = -\iu.
\label{eq:lowovertonespacing}
\end{equation}
A dividing line then exists between the first $2n$ modes and the rest of the modes, defined by a critical value of $\gamma^c_{n}$ which marks the $n$th bifurcation event. The first $2n$ modes limit to the spacing given by \eqref{eq:lowovertonespacing} while the higher unbifurcated modes limit to \eqref{eq:asymptoticspacing}.

By spectral similarity, an equivalent statement also follows for $\rho^c_{n,l}$ and $\lambda_{n,\ell,\rho}$. This will be the topic of the next section.

The bifurcation of the EM-SAdS spectrum seems to be important also since it bypasses the spacing predicted analytically in the asymptotic limit \cite{cardoso2004} for perturbations of the SAdS$_4$ spacetime with arbitrary spin weight. This analytic result, given by \eqref{eq:asymptoticspacing}, concerning the equal spacing for scalar, electromagnetic and gravitational perturbations of the SAdS black hole have been confirmed numerically as well \cite{berti2003, cardoso2003}.

As the overdamped modes populate the lowest lying overtones, we shall see when we solve the exact solution in the large black hole limit that modes with spacing \eqref{eq:asymptoticspacing} completely disappear, replaced by a spacing given by \eqref{eq:lowovertonespacing} all throughout the spectrum. What this tells us is that, for the EM-SAdS system, the large overtone limit and the large black hole limit are not commutative: depending on how you reach both limits, you would see a spectrum with spacing given by either \eqref{eq:asymptoticspacing} or \eqref{eq:lowovertonespacing}.

\section{Bifurcation master equation and a Feigenbaum-like constant}
\label{sec:master}

As we have discussed, the set of bifurcation points $\{\rho^c_{n, \ell}\}$ is very interesting. The symmetry \eqref{eq:symmetry} for pairs of quasinormal mode frequencies is broken and there is a corresponding jump in the frequency of the fundamental vibrational mode as well. These bifurcation points also indicate which parts of the spectrum limit to a spacing defined by \eqref{eq:asymptoticspacing} or \eqref{eq:lowovertonespacing}.

Here we shall derive a master equation that accurately approximates all values in this set, via a combination of analytic and numerical arguments.

Using the spectral similarity relation demonstrated in the previous section, we solve \eqref{eq:ReggeWheelerEMSAdSELBL} for critical values of $\{\gamma^c_n\}$ and relate
\begin{equation}
\rho^c_{n,l} \approx \sqrt{\dfrac{\gamma^c_n}{\ell(\ell+1)}}.
\label{eq:EMSAdSLcestimate}
\end{equation}
If \eqref{eq:EMSAdSLcestimate} is true, this would explain a feature in Fig. \ref{fig:bifurcation2}, related to the spacing of the inverse of $\rho^c_{n, \ell}$ for a given overtone and adjacent values of $\ell$. Consider the difference
\begin{equation}
\Delta \left(\rho^c_{n, \ell}\right)^{-1} = \left(\rho_{n,\ell}^c \right)^{-1} - \left(\rho_{n,\ell-1}^c \right)^{-1}.
\end{equation}
Equation \eqref{eq:EMSAdSgammaestimate} predicts that the spacing is equal in the eikonal limit,
\begin{equation}
\lim_{\ell \to \infty} \Delta \left(\rho^c_{n, \ell}\right)^{-1} = \sqrt{\dfrac{1}{\gamma^c_n}}
\end{equation}
The values of $\gamma^c_n$ may be numerically calculated via a binary search algorithm. The binary search algorithm uses the fact that below the critical value the $n$th overtone exists as a pair of modes satisfying \eqref{eq:symmetry}, while above the critical value the symmetry is broken and replaced by two pure overdamped modes. The algorithm terminates when at least 25 significant digits are recorded. A similar binary search algorithm may be implemented to numerically calculate $\rho^c_{n,\ell}$.

\begin{table}[t]
\begin{center}
\begin{tabular}{ccc}
$n$ & $\gamma^c_n$ & $\lambda^c_n$ \\
\hline \hline
0 & $1.151448404\times 10^{-1}$ & --1.365272160 $i$ \\
1 & $5.165673360\times 10^{-3}$ & --3.344965176 $i$ \\
2 & $1.739561715\times 10^{-4}$ & --5.340542424 $i$ \\
3 & $5.421427407\times 10^{-6}$ & --7.338560128 $i$ \\
4 & $1.625589807\times 10^{-7}$ & --9.337432770 $i$ \\
5 & $4.761150906\times 10^{-9}$ & --11.33670536 $i$ \\
6 & $1.372613841\times 10^{-10}$ & --13.33619716 $i$ \\
7 & $3.912424972\times 10^{-12}$ & --15.33582206 $i$ \\
8 & $1.105653757\times 10^{-13}$ & --17.33553383 $i$ \\
9 & $3.103722261\times 10^{-15}$ & --19.33530543 $i$ \\
10 & $8.665844274\times 10^{-17}$ & --21.33511998 $i$ \\
11 & $2.408920121\times 10^{-18}$ & --23.33496641 $i$ \\
12 & $6.671628545\times 10^{-20}$ & --25.33483715 $i$ \\
13 & $1.841985180\times 10^{-21}$ & --27.33472685 $i$ \\
14 & $5.071935664\times 10^{-23}$ & --29.33463162 $i$ \\
15 & $1.393316077\times 10^{-24}$ & --31.33454858 $i$ \\
\hline
\end{tabular}
\end{center}
\caption{Critical values of $\gamma^c_n$ of \eqref{eq:ReggeWheelerEMSAdSELBL} such that the $n$th overtone reaches the negative imaginary axis at $\lambda^c_n$.}
\label{tab:gammacn}
\end{table}

\begin{figure}[b]
\begin{center}
\includegraphics[width=0.45\textwidth]{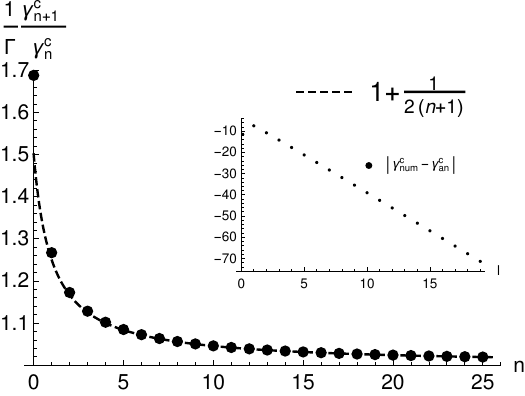}
\end{center}
\caption{Plot of $ \dfrac{1}{\Gamma} \dfrac{\gamma^c_{n+1}}{\gamma^c_n}$ and the fitting $\left( 1 + \dfrac{1}{2(n+1)}\right)$ which motivates \eqref{eq:EMSAdSgammaestimate}. The inset gives the absolute difference between the numerical values found via binary search and the analytical estimate \eqref{eq:gammacnestimate}, showing exponential convergence.}
\label{fig:ARatios}
\end{figure}

The first 16 values are given in Table \ref{tab:gammacn}, where we only show the first 10 significant digits and have also included the value of the $n$th overtone when it hits the negative imaginary axis. So as not to overload our notation, we define
\begin{equation}
\lambda^c_{n} \equiv \lambda_{\gamma^c_{n}, n}.
\end{equation}
The data in Table \ref{tab:gammacn} are interesting, since they numerically show that
\begin{equation}
\lim_{n\to\infty} \left( \lambda^c_{n+1} - \lambda^c_{n} \right) = -2 \iu
\end{equation}
and
\begin{equation}
\lim_{n\to\infty} \dfrac{\gamma^c_{n+1}}{\gamma^c_n} = \Gamma, \qquad \Gamma = \exp \left( -\dfrac{2 \pi}{\sqrt{3}}\right).
\end{equation}
This last expression is reminiscent of a Feigenbaum constant, since it is equivalent to
\begin{equation}
\lim_{n \to \infty} \dfrac{\gamma_{n+1}^c-\gamma_n^c}{\gamma_{n}^c - \gamma_{n-1}^c} = \Gamma.
\end{equation}
In fact, a more accurate expression is given by
\begin{equation}
\dfrac{\gamma^c_{n+1}}{\gamma^c_n} \approx \left(1 + \dfrac{1}{2(n+1)} \right) \Gamma.
\label{eq:EMSAdSgammaestimate}
\end{equation}
This expression is numerically motivated by Fig. \ref{fig:ARatios}, where we have plotted $\dfrac{1}{\Gamma} \dfrac{\gamma^c_{n+1}}{\gamma^c_n} $ and matched it with
\begin{equation}
\dfrac{1}{\Gamma} \dfrac{\gamma^c_{n+1}}{\gamma^c_n} \approx \left(1 + \dfrac{1}{2(n+1)} \right).
\end{equation}
In fact, once the Feigenbaum-like constant was noticed, the expression \eqref{eq:EMSAdSgammaestimate} became vital in speeding up the binomial search algorithm, since it may be used to define both the initial center and the interval of the search to great precision.

Thus
\begin{equation}
\gamma^c_n \approx \gamma_0^c \left[ \prod_{k=1}^n \left( 1+ \dfrac{1}{2k} \right) \right] \Gamma
\end{equation}
or we may fit
\begin{equation}
\gamma^c_n = (1 + \epsilon) \gamma_0^c \dfrac{(2n+1)!!}{2^n n!}\Gamma
\label{eq:gammacnestimate}
\end{equation}
where the constant $\epsilon$ accumulates all the deviations from \eqref{eq:EMSAdSgammaestimate}. This constant is numerically consistent with
\begin{equation}
\epsilon = 8.84744 \times 10^{-5}
\end{equation}
The agreement between the numerically calculated values of $\gamma^c_n$ and \eqref{eq:gammacnestimate} is excellent, as shown in the inset of Fig. \ref{fig:ARatios} which plots the absolute difference between the two. The inset shows exponential convergence between the numerical values and the analytic expression.

\begin{figure}[t]
\begin{centering}
\begin{tabular}{c}
  \includegraphics[width=0.45\textwidth]{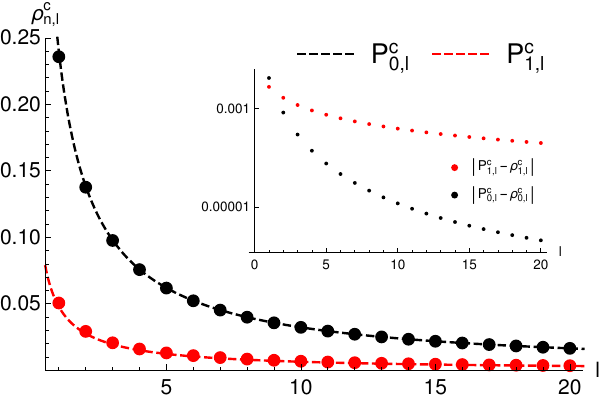}
\end{tabular}
\caption{The critical values of $\rho^c_{n,l}$ defined to be the value of $\rho$ for harmonic mode index $\ell$ so that the $n$th overtone of \eqref{eq:ReggeWheelerEMSAdS} reaches the negative imaginary axis, for $n = 0$ and $n = 1$. Plotted also is the estimate \eqref{eq:EMSAdSLcestimate}. The inset gives the difference between the numerical value and analytical estimate, showing the expected power-law falloff with a form given by \eqref{eq:falloff}.}
\label{fig:Ldeltarh}
\end{centering}
\end{figure}

Finally, we arrive at our master equation
\begin{equation}
P^c_{n,\ell} = \sqrt{(1 + \epsilon) \gamma_0^c \dfrac{(2n+1)!!}{2^n n!\ell(\ell+1)}\Gamma},
\label{eq:masterequation}
\end{equation}
where
\begin{equation}
\rho^c_{n,\ell} \approx P^c_{n,\ell}.
\end{equation}
We show how closely our analytic expression matches the numerically solved values of the bifurcation points in Fig.~\ref{fig:Ldeltarh}. The inset shows a power-law falloff of the absolute difference between the numerical value and analytical estimate. This is expected, since \eqref{eq:EMSAdSLcestimate} is a zeroth order approximation of \eqref{eq:perturbed}. The form of the perturbing parameter implies that the first correction term is given by
\begin{equation}
\left|\rho^c_{n,l} - \sqrt{\dfrac{\gamma_n^c}{\ell(\ell+1)}} \right|  \sim \dfrac{1}{\ell(\ell +1)}.
\label{eq:falloff}
\end{equation}

\section{Exact solutions in the large black hole limit}
\label{sec:LBHL}

In the large black hole limit $\rho \to 0$, the ODE eigenvalue problem \eqref{eq:ReggeWheelerEMSAdS} reduces to
\begin{equation}
(u^3 - 1) \deri{^2\phi_{\lambda}(u)}{u^2} + 3 (u^2 - \iu \lambda ) \deri{\phi_{\lambda}(u)}{u} = 0
\label{eq:exactODELBL}
\end{equation}
One of the solutions is a constant. Since the above equation may be written as
\begin{equation}
\deri{}{u} \left( (u^3 - 1) \deri{\phi_{\lambda}(u)}{u} - 3 \iu \lambda \phi_{\lambda}(u) \right) = 0,
\end{equation}
the second linearly independent solution satisfies the separable equation,
\begin{equation}
(u^3 - 1) \deri{\phi_{\lambda}(u)}{u} - 3 \iu \lambda \phi_{\lambda}(u) = 0,
\end{equation}
whose solution is of the form
\begin{equation}
\phi_{\lambda}(u) = A g(u)^{\iu \lambda /2} .
\end{equation}
where
\begin{equation}
g(u) = \dfrac{(1-u)^2}{1 + u + u^2} \exp \left( - 2 \sqrt{3}  \arctan \left[ \dfrac{1 + 2u}{\sqrt{3}} \right] \right).
\end{equation}
Since the above solution has the asymptotic behavior
\begin{equation}
\phi_{\lambda}(u) \sim (1 - u)^{\iu \lambda},
\end{equation}
and we recall that the Frobenius expansion must be regular around $u = 1$, this gives us the eigenvalues
\begin{equation}
\lambda = 0, - \iu, - 2 \iu, - 3 \iu, \dots, - n \iu, \dots \qquad n \in \mathbb{Z}.
\end{equation}
This spectrum matches the spacing \eqref{eq:lowovertonespacing}; no overtones with spacing \eqref{eq:asymptoticspacing} exist, even in the high overtone limit.

The zero mode corresponds to a constant eigenfunction,
\begin{equation}
\phi_0(u) = A.
\end{equation}
The Dirichlet boundary condition \eqref{eq:TIRBC} corresponding to a total internal reflection at spatial infinity does not support the zero mode, since the corresponding eigenfunction vanishes. However, other boundary conditions may support the zero mode \cite{wang2021, wang2021a}, whenever the boundary conditions permit a constant nonzero solution.

Thus, the general solution of the EM SAdS in the large black hole limit is of the form,
\begin{equation}
\phi_n(u) = A g(u)^{n/2} + B, \qquad n \in \mathbb{Z}
\label{eq:mostgensol}
\end{equation}

In deriving the above general solution, we have so far only imposed the regularity of the solution after scaling out the asymptotic behavior of the causal solution around the black hole boundary. Thus, different boundary conditions at the spatial infinity would approach this universal spectrum in the large black hole limit. 

This has been observed numerically before, at least comparing Dirichlet and Robin boundary conditions for gravitational perturbations \cite{moss2002}. But the exact solution in the large black hole limit enables the first (to the best of our knowledge) analytic treatment both confirming and explaining this numerical observation, at least for electromagnetic perturbations.

When we consider the dimensionful eigenvalues, $\bar{\omega}$ correspond to infinitely damped modes. These seem to play an important role in theories for quantum gravity \cite{hod1998, dreyer2004, cardoso2004}.

\subsection{Comment on the satisfaction of the causal boundary conditions}

The general solution we have derived corresponds to a case where the indicial exponents are integer separated. While it is common that the subdominant solution has a logarithmic tail, here the subdominant solution is just a constant function. For the total internal reflection boundary condition,
\begin{equation}
\phi_n(u) = A \left( 1 - \Gamma^{-1/4} g(u)^{n/2} \right),
\label{eq:TIRsol}
\end{equation}
where we have used the Feigenbaum-like constant that appeared in a previous section. At this point, we are reminded of the subtleties concerning the algebraically special modes of asymptotically flat spacetimes \cite{chandrasekhar1984,maassenvandenbrink2000}. Naively, the solutions satisfying the total internal reflecting boundary condition seem to imply that the overdamped modes are a mixture of causal and acausal waves around the black hole boundary. That is, the overdamped modes are not real quasinormal modes -- very similar to what occurs in asymptotically flat spacetimes \cite{maassenvandenbrink2000,cook2016}. \footnote{We recall the following for the asymptotically flat spacetimes: after a rescaling similar to \eqref{eq:rescaling} and at the algebraically special frequencies, both local solutions around the black hole horizon corresponding to ingoing and outgoing solutions are analytic, owing to a special disappearance of the logarithm that usually accompanies the subdominant solution. This means that the analyticity of the global solution does not guarantee that the eigensolution at the algebraically special frequency satisfies the quasinormal mode boundary conditions around the black hole horizon.}

However, explicitly determining whether the solution \eqref{eq:TIRsol} satisfies the causal boundary condition \eqref{eq:bccausal} is difficult because the dimensionful frequency $\bar{\omega}$ diverges in the large black hole limit. The plane wave prescription around the black hole boundary thus breaks down.

For finite but large black holes, the indicial exponents of the overdamped eigenfunctions are no longer integer separated. If the overdamped eigenfunctions are a mixture of causal and acausal parts, we expect that these solutions are not smooth around the black hole boundary.

Since the acausal indicial exponent corresponding to each overdamped mode is some positive noninteger value, it is straightforward to try and look for divergent behavior around the black hole boundary in the derivatives of an overdamped eigenfunction. We have not observed this for the eigenfunctions we have numerically calculated -- they all seem perfectly smooth around the black hole boundary -- suggesting that they are bona fide quasinormal modes.

\subsection{Expansion around the large black hole limit: Overdamped modes}

We shall now attempt to quantify the insensitivity of the spectrum to the boundary conditions at spatial infinity around the large black hole limit. In our discussion from Sec. \ref{sec:selfsim}, the ODE eigenvalue problem reduces to
\begin{equation}
\left( u^3 - 1 \right) \deri{^2 \phi(u)}{u^2} + 3 \left( u^2 - \iu \lambda \right) \deri{\phi(u)}{u} + \gamma \phi(u) = 0
\end{equation}
in the large black hole limit. We start our analysis with this ODE. Neither eigenfunction is convenient for use, because of a complicated recurrence relation with its derivatives.

We thus rescale the eigenfunctions so that the general solution has a useful form,
\begin{equation}
\phi_{\eta,l}(u) = \left( \dfrac{\sqrt{g(u)}}{1-u} \right)^{\eta} \psi_{\eta,l}(u).
\end{equation}
For simplicity, we shall also be applying a transformation on the eigenvalues,
\begin{equation}
\lambda = - \iu \eta_{\gamma}
\end{equation}
so that the expansion of $\eta_{\gamma}$ around $\gamma = 0$ has the form
\begin{equation}
\eta_\gamma = n + \sum_{k=0}^\infty A_k(n) \gamma^k.
\label{eq:gammaexpand}
\end{equation}
The resulting differential equation is of the form
\begin{multline}
\left( u^3 - 1 \right) \deri{^2 \psi_{\eta}(u)}{u^2}
+ (\eta_\gamma - 2 \eta_\gamma u + (3- 2 \eta_\gamma) u^2 ) \deri{\psi_{\eta}(u)}{u} \\
+ ( \gamma + (2 \eta_\gamma -1 ) \eta_\gamma + ( \eta_\gamma - 2) \eta_\gamma u ) \psi_\eta(u) = 0,
\end{multline}
where there are again two linearly independent solutions in the limit $\gamma \to 0$,
\begin{equation}
\psi_{\eta_0}(u) = A \psi_{n,1}(u) + B \psi_{n,2}(u)
\end{equation}
where
\begin{equation}
\psi_{n,1}(u) = \left( \dfrac{1-u}{\sqrt{g(u)}} \right)^{n}, \qquad \psi_{n,2}(u) = (1-u)^n.
\end{equation}
This is more convenient, since now $(1-u)^n$ forms a complete basis, whose derivatives satisfy a simple recurrence relation.

One of these constants of the general solution will be determined by the boundary conditions at spatial infinity. We shall keep our analysis very general. Suppose some boundary condition at infinity imposes an eigenfunction of the form
\begin{equation}
\psi_{\eta_0}(u) = A \left( \psi_{n,1}(u) + \xi \psi_{n,2}(u) \right).
\label{eq:psieta1}
\end{equation}
For the total internal reflecting boundary condition, we have
\begin{equation}
\xi = - \Gamma^{-1/4},
\end{equation}
where curiously $\Gamma$ is the Feigenbaum-like constant we have identified earlier. 

For finite $\gamma$, we do not expect two regular solutions around the black hole boundary. Thus, we choose
\begin{equation}
\psi_{\eta_{\gamma}}(u) = A \left( \psi_{n,1}(u) + \sum_{k=1}^\infty \gamma^k \psi_{\eta,k}(u) \right).
\label{eq:psieta2}
\end{equation}
Now, we express both \eqref{eq:psieta1} and \eqref{eq:psieta2} as
\begin{equation}
\psi_{\eta_0}(u) = A \sum_{k=0}^\infty c_k (1-u)^k
\end{equation}
and
\begin{equation}
\psi_{\eta_{\gamma}}(u) = A \sum_{k=0}^\infty d_k(\gamma) (1-u)^k
\label{eq:psietagen}
\end{equation}
respectively, where
\begin{equation}
\begin{matrix*}[l]
c_k = \dfrac{(-1)^k}{k!} \deri{^k}{u^k} \psi_{\eta_0}(u=1), \vspace{3pt}\\
d_k(\gamma) = \dfrac{(-1)^k}{k!} \deri{^k}{u^k} \psi_{\eta_{\gamma}}(u=1).
\end{matrix*}
\end{equation}
We note that
\begin{equation}
c_k = \dfrac{(-1)^k}{k!} \deri{^k}{u^k} \psi_{n,1}(u=1) + (-1)^k \xi \delta_{k,n}
\end{equation}
and
\begin{equation}
d_k(0) = \dfrac{(-1)^k}{k!} \deri{^k}{u^k} \psi_{n,1}(u=1),
\end{equation}
where $\delta_{k,n}$ is the Kronecker delta
\begin{equation}
\delta_{k,n} = \left\lbrace
\begin{matrix}
0, \qquad k \neq n \\
1, \qquad k = n
\end{matrix}
\right. .
\end{equation}
Consider the recurrence relation satisfied by $c_k$ and $d_k(\gamma)$ for $\gamma = 0$ and finite $\gamma$,
\begin{equation}
(1 + k - \eta_0)c_{k-1} + 3(k-\eta_0) c_k - 3(k+1)c_{k+1} = 0
\end{equation}
and 
\begin{multline}
(1 - k + \eta_\gamma) d_{k-1}(\gamma) + \left( \dfrac{\gamma}{1+k - \eta_\gamma} + 3(k - \eta_\gamma) \right) d_k(\gamma) \\ 
- 3(n+1) d_{k+1}(\gamma) = 0.
\end{multline}
Note the extra term for the recurrence relation for $d_k(\gamma)$, which remains finite as you approach $\gamma \to 0$. That is, if we choose $k = n - 1$ and use \eqref{eq:gammaexpand}, we have for the both of them
\begin{equation}
c_{n-1} + n c_{n} = 0
\end{equation}
and
\begin{equation}
\left( -\dfrac{1}{3 A_1(n)} + 1 \right) d_{n-1}(0) + n d_{n}(0) + \mathit{O}(\gamma) = 0.
\end{equation}
If we assume that $\psi_{\eta_\gamma}(u)$ is continuously related to $\psi_{\eta_0}(u)$ in the limit $\gamma \to 0$, these two expressions must match in that same limit. Thus, we arrive at an expression for the first expansion coefficient of $\eta_\gamma$,
\begin{equation}
A_1(n) = (-1)^{n}\dfrac{d_{n-1}(0)}{3 n \xi} =  \dfrac{(-1)^{n}}{3 \xi n!} \deri{^{n-1}}{u^{n-1}} \psi_{n,1}(u=1).
\label{eq:A1(n)}
\end{equation}
That is,
\begin{equation}
\eta_\gamma \approx n + \gamma A_1(n)
\end{equation}
or
\begin{equation}
\lambda = - \iu n - \iu \gamma A_1(n).
\end{equation}
We note that this expression works for all overtone numbers except for $n = 0$, and for any boundary condition imposed at spatial infinity that results in a quasinormal mode spectrum. Thus, black hole spacetimes cannot vibrate in the large black hole limit independent of the boundary condition at spatial infinity.

\subsection{Expansion around the large black hole limit: The zero mode}

Calculating the expansion around the zero mode is less straightforward. Equation \eqref{eq:psieta1} is no longer valid, since the eigenfunction of the zero mode is simply a constant function,
\begin{equation}
\psi_{\eta_0}(u) = A.
\end{equation}
We shall use this fact, and assume that for \eqref{eq:psietagen},
\begin{equation}
\lim_{\gamma \to 0} d_0(\gamma) \sim 1, \qquad \lim_{\gamma \to 0} d_1 (\gamma) \sim \gamma.
\end{equation}
Then, to determine, $A_1(0)$ from
\begin{equation}
\eta_\gamma = \sum_{k=0}^\infty A_1(0) \gamma^k
\end{equation}
we use a combination of the recurrence relation
\begin{equation}
(\gamma + 3 \eta_\gamma (\eta_\gamma - 1) ) d_0(\gamma) + 3 (\eta_\gamma - 1) d_1(\gamma) = 0
\label{eq:zeromoderecurrence}
\end{equation}
and the boundary condition at spatial infinity that supports a constant solution. As a sample calculation, let us consider the Robin boundary condition corresponding to a conservation of energy condition at spatial infinity \cite{wang2021,wang2021a},
\begin{equation}
2 \deri{\psi_{\eta_\gamma}(0)}{u} - \eta_\gamma \psi_{\eta_\gamma}(0) = 0.
\end{equation}
The boundary condition leads us to
\begin{equation}
d_1(\gamma) = - \dfrac{\eta_\gamma}{2} d_0(\gamma).
\end{equation}
We folded this back in to \eqref{eq:zeromoderecurrence} to yield
\begin{equation}
(\gamma + \dfrac{3}{2} \eta_\gamma (\eta_\gamma - 1)) d_0(\gamma) = 0.
\end{equation}
For the constant solution to be supported in the limit $\gamma \to 0$, the above equation must be true while $d_0(\gamma) \neq 0$. That is,
\begin{equation}
\gamma + \dfrac{3}{2} \eta_\gamma (\eta_\gamma - 1) = 0, \qquad \eta_\gamma = \sum_{k=1}^\infty A_k(0) \gamma^k
\end{equation}
This yields
\begin{equation}
A_1(0) = \dfrac{2}{3}.
\end{equation}

\subsection{Testing the expansion around the large black hole limit}

We test our expansion around the large black hole limit. We shall be using two boundary conditions. First, the Dirichlet boundary conditions we have used in the bulk of this paper,
\begin{equation}
\psi_{\eta_\gamma}(0) = 0, \qquad \xi_{D} = - \Gamma^{-1/4}
\end{equation}
and Robin boundary conditions,
\begin{equation}
2 \deri{\psi_{\eta_\gamma}(0)}{u} - \eta_\gamma \psi_{\eta_\gamma}(0) = 0, \qquad \xi_R = \Gamma^{-1/4}.
\label{eq:kappaD}
\end{equation}
Both of these conditions conserve energy at spatial infinity.

Since
\begin{equation}
\xi_R = - \xi_D,
\end{equation}
then
\begin{equation}
A_{1,R}(n) = - A_{1,D}(n).
\end{equation}

We define the finite differences,
\begin{equation}
\Delta_D(\gamma) = \dfrac{\eta_{\gamma,D} - n}{\gamma}, \qquad \Delta_R(\gamma) = \dfrac{\eta_{\gamma,R} - n}{\gamma},
\label{eq:finitedifference}
\end{equation}
where $\eta_{\gamma,D}$ and $\eta_{\gamma,R}$ is a quasinormal mode calculated using either the Dirichlet or Robin boundary conditions respectively, and compare
\begin{equation}
A_{1,R}(n) \approx -\Delta_D(\gamma) \approx \Delta_R(\gamma)
\label{eq:A1(n)approx}
\end{equation}
for some small $\gamma$. Note that the Dirichlet boundary conditions does not support the zero mode.

In Table \ref{tab:expansion}, we do this comparison for $\gamma = 10^{-12}$. We note there is excellent agreement between the analytic expression and the expansion around the large black hole limit for either boundary condition, since the table shows a range of matching from all digits shown at $n=0$ and six digits at $n=8$ for both boundary conditions. For ease of comparison, we show the exact expression for $A_{1,R}(n)$ and also show its decimal expansion.
\begin{table}[t]
\begin{center}
\begin{tabular}{ccccc}

$n$ & Exact & Decimal & $-\Delta_D(\gamma)$ & $\Delta_R(\gamma)$ \\
\hline \hline \vspace{-5pt} \\
0 & $ \dfrac{2}{3} $ & 0.666666667 & \dots & 0.666666667 \vspace{2pt} \\
1 & $-\dfrac{1}{\sqrt{3}} \Gamma^{-1/4}$  & --1.429884308 & --1.429884308 & --1.429884308 \vspace{2pt}\\
2 & $\Gamma^{-1/2}$ & 6.133707406 & 6.133707406 & 6.133707406 \vspace{2pt}\\
3 & $-\dfrac{7}{2\sqrt{3}} \Gamma^{-3/4}$ & --30.69672190 & --30.69672191 & --30.69672190 \vspace{2pt}\\
4 & $\dfrac{13}{3} \Gamma^{-1}$ & 163.0302550 & 163.0302550 & 163.0302551 \vspace{2pt}\\
5 & $ -\dfrac{133}{8\sqrt{3}} \Gamma^{-5/4}$ & --894.3523748 & --894.3523762 & --894.3523734 \vspace{2pt}\\
6 & $ \dfrac{217}{10} \Gamma^{-3/2} $ & 5007.591567 & 5007.591523 & 5007.591610 \vspace{2pt}\\
7 & $-\dfrac{20683}{240\sqrt{3}} \Gamma^{-7/4} $ & --28436.25211 & --28436.25352 & --28436.25070 \vspace{2pt}\\
8 & $\dfrac{1729}{15} \Gamma^{-2} $ & 163153.3347 & 163153.2881 & 163153.3814 \vspace{2pt}\\
\hline
\end{tabular}
\end{center}
\caption{Comparison between the analytic and numerical first order expansions around the black hole limit for Dirichlet and Robin boundary conditions. Exact and decimal expressions were calculated using \eqref{eq:A1(n)} and \eqref{eq:kappaD}, and numerical expressions were calculated using the finite differences \eqref{eq:finitedifference} with $\gamma = 10^{-12}$. There is excellent agreement between the analytic and numerical expressions, since the table shows a range of matching from all digits shown at $n=0$ and six digits at $n=8$ for both boundary conditions.}
\label{tab:expansion}
\end{table}

\section{Conclusion}

In this study, we have extensively explored the nature of electromagnetic perturbations of Schwarzschild--anti--de Sitter black holes in the large black hole limit. We have presented three major results: (1) a novel symmetry of the quasinormal mode spectrum which we call spectral similarity, (2) a master equation which describes at what values of the dimensionless constant $\rho$ the bifurcations in the spectrum occur and (3) exact solutions of both the spectrum and the eigenfunctions in the large black hole limit, as well as the first order expansion of the spectrum around the same limit.

Between these three results, we expound on prior studies on the EM-SAdS system, as well as answer open questions found in the literature. 

Spectral similarity is sufficient in explaining the $\ell$ independence of the spectrum in the large black hole limit. We have also shown that the level spacing of the spectrum in the same limit is a separate effect, attributed to the properties of a related eigenvalue problem which results when simultaneously reaching the large black hole limit and the eikonal limit.

We have expounded on the bifurcations in the spectrum \cite{wang2021,wang2021a}, detailing the various sudden changes in the qualities of the EM-SAdS system around the bifurcation. Apart from a change in the number of pairs of modes satisfying the symmetry \eqref{eq:symmetry} and a jump in the fundamental ``note" of electromagnetic perturbations, we have shown that the bifurcation marks a shift in which of two asymptotic spacings sets of quasinormal modes reach in the large black hole limit: either given by \eqref{eq:asymptoticspacing} or \eqref{eq:lowovertonespacing} for the unbifurcated and bifurcated modes respectively.

The master equation describing when the bifurcations occur is itself very interesting because of the emergence of a Feigenbaum-like constant which describes a geometric speedup between consecutive bifurcations. For the related eigenvalue problem, a part of the master equation predicts the bifurcations with exponential convergence. For the full master equation, there is a power-law convergence, fully explained by our discussion.

With the exact solution, we show that the spacetime cannot vibrate in the large black hole limit and, at least for electromagnetic perturbations, explain the insensitivity of the spectrum to the boundary conditions at spatial infinity. We quantify the effect of the boundary conditions, which only appear at first order around the large black hole limit.

Some of our results can be expected to have interesting implications for the AdS/CFT correspondence, but they are also of intrinsic interest insofar as they help to clarify some of the peculiar properties of electromagnetic perturbations in asymptotically AdS spacetimes that have fascinated the community. We leave to future work extensions to scalar and gravitational perturbations and to the Kerr-AdS spacetime.

The code we have used, which we call \texttt{SpectralBP}, is publicly available and may be found at \url{https://github.com/slashdotfield/SpectralBP}. Details of its implementation in a \texttt{Mathematica} package may be found at \cite{fortuna2021}.

\section*{ACKNOWLEDGEMENTS}
We are grateful to J. Celestial for providing insightful comments about the exact solutions in the large black hole limit, and to E. Poisson for encouraging us to think hard about the physical meaning of our solutions and to reconsider the terminology we used in an earlier version of this paper. S.F. is supported by the Department of Science and Technology Advanced Science and Technology Human Resources Development Program - National Science Consortium. This research is supported by the University of the Philippines Diliman Office of the Vice Chancellor for Research and Development through Project No.~191937 ORG.

\bibliographystyle{apsrev4-1}
\bibliography{bibfile}
\onecolumngrid 

\end{document}